\documentclass[12pt]{article}
\usepackage{amsmath,amsthm, amssymb}

\newcommand{\beq}{\begin{equation}}
\newcommand{\eeq}{\end{equation}}
\newcommand{\ba}{\begin{array}}
\newcommand{\ea}{\end{array}}

\newcommand{\bea}{\begin{eqnarray}}
\newcommand{\eea}{\end{eqnarray}}
\newcommand{\bma}{\begin{matrix}}
\newcommand{\ema}{\end{matrix}}
\newcommand{\bpm}{\begin{pmatrix}}
\newcommand{\epm}{\end{pmatrix}}
\newcommand{\nn}{\nonumber}
\newcommand{\half}{\textstyle{\frac{1}{2}}}

\newcommand{\qstar}{q_{12}^{\ast}}
\newcommand{\qq}{\vec q ^{\,2}}

\newcommand{\dvec}[1]{\stackrel{\rightarrow}{#1}}
\newcommand{\sx}{\sqrt6}
\newcommand{\Dh}{\widehat D}
\newcommand{\p}{\partial}
\newcommand{\hf}{{\hat5}}
\newcommand{\effh}{{e_5^\hf}}
\newcommand{\efhf}{{e_\hf^5}}
\newcommand{\Qij}{Q_i{}^j}
\newcommand{\Uij}{U_i{}^j}

\newcommand{\pr}{\prime}
\newcommand{\ov}{\overline}
\newcommand{\wh}{\widehat}
\newcommand{\wt}{\widetilde}
\newcommand{\Psitil}{\wt \Psi}
\newcommand{\Psibar}{\ov \Psi}
\newcommand{\Eta}{\mathcal H}
\newcommand{\Etatil}{\wt \Eta}

\newcommand{\psibar}{\ov \psi}
\newcommand{\etabar}{\ov \eta}
\newcommand{\sibar}{\ov \sigma}

\newcommand{\ep}{\varepsilon}
\newcommand{\eps}{\epsilon}
\newcommand{\al}{\alpha}
\newcommand{\aone}{\alpha_1}
\newcommand{\dz}{\delta(z)}
\newcommand{\la}{\lambda}
\newcommand{\da}{\delta}
\newcommand{\om}{\omega}
\newcommand{\Ga}{\Gamma}
\newcommand{\Si}{\Sigma}
\newcommand{\si}{\sigma}

\title{\bf Supersymmetric branes\\ with (almost) arbitrary tensions}
\author{\\[1cm]
 \large{\bf Jonathan A.~Bagger}~\thanks{bagger@jhu.edu} 
 \quad \it{and} \quad    
 \large{\bf Dmitry V.~Belyaev}~\thanks{belyaev@pha.jhu.edu}\\[1cm]
 \it Department of Physics and Astronomy\\ 
 \it The Johns Hopkins University\\
 \it 3400 North Charles Street\\
 \it Baltimore, MD 21218, USA}
\date{}

\begin{document}

\numberwithin{equation}{section}

\maketitle
\thispagestyle{empty}

\begin{abstract}
We present a supersymmetric version of the two-brane Randall-Sundrum 
scenario, with arbitrary brane tensions $T_1$ and $T_2$, subject to the bound 
$|T_{1,2}|\leq\sqrt{-6\Lambda_5}$, where $\Lambda_5<0$ is the bulk 
cosmological constant.  Dimensional reduction gives $N=1$, $D=4$ 
supergravity, with cosmological constant $\Lambda_4$ in the range
$\half\Lambda_5\leq\Lambda_4\leq 0$.  The case 
with $\Lambda_4=0$ requires $T_1=-T_2=\sqrt{-6\Lambda_5}$.  This work 
unifies and generalizes previous approaches to the 
supersymmetric Randall-Sundrum scenario.  It also shows that the 
Randall-Sundrum fine-tuning is not a consequence of supersymmetry.
\end{abstract}

\newpage

\setcounter{page}{1}
\section{Introduction}

During the past few years, codimension-one branes have been the subject 
of intense activity.  Much of this work was sparked by Randall and 
Sundrum, who showed how codimension-one branes can solve the gauge 
hierarchy problem \cite{rs}.  In this paper we consider supersymmetric 
extensions of the original five-dimensional Randall-Sundrum scenario.  
We compactify the fifth dimension on an $S^1/ \mathbb{Z}_2$ orbifold, 
and place codimension-one branes at the orbifold fixed points.  We 
require odd bosonic fields to be continuous across the branes, but we
let odd fermionic fields jump in a way that is consistent with their 
five-dimensional equations of motion.

In previous work on this subject, the brane tensions were tuned to be 
equal and opposite, and equal in magnitude to the five-dimensional bulk 
cosmological constant, appropriately normalized \cite{abn}--\cite{bkvp}.
In this paper we relax 
this condition and allow arbitrary brane tensions.  We present a 
bulk-plus-brane action and find the conditions under which it is locally 
supersymmetric.  Our results imply that the Randall-Sundrum fine tuning is 
not a consequence of supersymmetry:  it must be imposed by hand to 
obtain a flat effective four-dimensional theory.\footnote{Zucker came
to a similar conclusion using the off-shell formalism \cite{zuck},
but he was unable to find the Killing spinor that describes the
unbroken supersymmetry.}
More generally, our construction allows the locally supersymmetric
five-dimensional theory to have effective four-dimensional theory
with a negative or zero cosmological constant.

Moreover, in previous work, the bulk gravitino mass was taken to be 
either even \cite{abn} or odd \cite{gp, flp, bkvp} under the 
$\mathbb{Z}_2$ parity.  The approach presented here allows the results 
to be continuously connected -- even in the case of arbitrary brane 
tensions.  Our results show that there is no conceptual difference -- at least in
the absence of matter -- between the two cases previously considered
in the literature.

The paper is organized as follows.  In section 2 we present the 
supersymmetric bulk-plus-brane action, together with the corresponding 
supersymmetry transformations.  We find that supersymmetry requires the 
brane tensions to have magnitudes less than or equal to the bulk 
cosmological constant, appropriately normalized.  In section 3 we 
compute the low energy effective action.  We keep the radion field fixed 
and derive the effective action for the four-dimensional supergravity 
multiplet.  We show that the reduction leads to a locally supersymmetric 
theory with a negative or zero cosmological constant.  We summarize our 
conventions and present details of our calculation in Appendices A and B.

\section{Locally supersymmetric bulk-plus-brane system}

\subsection{Bulk-plus-brane action}

In this section we present our five dimensional bulk-plus-brane action.  We
start with pure $N=2,\ D=5$ supergravity, with cosmological constant $\Lambda_5
= -6\la^2$.  The cosmological constant arises from gauging a $U(1)$ subgroup of
the $SU(2)$ automorphism group, determined by a unit vector of real parameters
$\vec q = (q_1,q_2,q_3).$  The action and supersymmetry transformations
\cite{hist} are given by \footnote{The supersymmetry is $N=2$ because it
corresponds to a supersymmetry algebra with two independent supercharges, each
belonging to the smallest spinor representation (pseudoreal, of real dimension
4) of the Lorentz group $SO(1,4)$ \cite{stra}. However, because this is the
minimal algebra in $D=5$, this theory is sometimes called $N=1$.}
\bea
\label{bulk}
S_{\rm bulk} = \int d^5\!x e_5 
\Big\{
-\frac{1}{2} R + 6\la^2 
+\frac{i}{2}\Psitil_M^i \Ga^{MNK} D_N \Psi_{Ki}
-\frac{3}{2} \la\, \vec q\cdot \vec\si_i{}^j \Psitil_M^i \Si^{MN}  \Psi_{Nj} 
\nn\\
-\frac{1}{4}F_{MN}F^{MN} 
- i\frac{\sqrt6}{16}F_{MN}
\left(
2\Psitil^{Mi} \Psi_i^N + \Psitil_P^i \Ga^{MNPQ} \Psi_{Qi} 
\right) 
\nn\\
-\frac{1}{6\sqrt6}\eps^{MNPQK} F_{MN}F_{PQ}B_K
+\frac{\sqrt6}{4} \la\, \vec q\cdot \vec\si_i{}^j
B_N\Psitil_M^i\Ga^{MNK}\Psi_{Kj}
\Big\}
\eea
and
\bea
\label{bulksusy}
\da e_M^A &=& i \Etatil^i\Ga^A\Psi_{Mi} \nn\\
\da B_M &=& i\frac{\sqrt6}{2}\Psitil_M^i\Eta_i \nn\\
\da \Psi_{Mi} &=& 2 \big(
D_M\Eta_i - i\frac{\sqrt6}{2}\la\, \vec q\cdot \vec\si_i{}^j B_M\,\Eta_j \big)
+i\la\, \vec q\cdot \vec\si_i{}^j \,\Ga_M \,\Eta_j \nn\\
&& +\frac{1}{2\sqrt6} \left(
\Ga_{MNK} - 4g_{MK}\Ga_N \right) F^{NK}\Eta_i ,
\eea
where we drop all three- and four-Fermi terms, and the spinors are symplectic
Majorana (see Appendix A).  For the case at hand, we write a symplectic
Majorana spinor $\Psi_i$ as follows,
\bea
\Psi_1 = -\Psi^2 = \binom{\psi_{1\alpha}}{\psibar_2^{\dot\alpha}}, \quad
\Psi_2 =  \Psi^1 = \binom{-\psi_{2\alpha}}{\psibar_1^{\dot\alpha}} ,
\eea
where $\psi_{1\alpha}$ and $\psi_{2\alpha}$ are two-component Weyl spinors.

We take the fifth dimension to span the orbifold $\mathbb{R}/\mathbb{Z}_2$.
We work on the covering space and require that the action and supersymmetry
transformations be invariant under reflection $z \rightarrow -z$, where $z=x^5$
is the coordinate in the fifth dimension.  (We consider $S^1/\mathbb{Z}_2$
later in this section.)  We assume the action is even under reflection.  We
then choose $e_m^a,$ $\eta_1$ and $\lambda$ to be even, which fixes the
remaining parity assignments:
\bea
\label{paras}
\bma
\text{even}:&\p_m&e_m^a&\effh&B_5&
\eta_1&\psi_{m1}&\psi_{52}&q_{1,2}&{\la} \\
\text{ odd}:&\p_5&e_m^{\hf} & e_5^a&B_m&
\eta_2&\psi_{m2}&\psi_{51}&q_3 ,&{}
\ema
\eea
where $(\eta_1, \eta_2)$, $(\psi_{m1}, \psi_{m2})$ and $(\psi_{51}, \psi_{52})$
are the two-component spinors in $\Eta_i$, $\Psi_{mi}$ and $\Psi_{5i}$,
respectively.

The orbifold fixed point at $z=0$ can be viewed as a 3-brane, $\Si$,
parametrized by the coordinates $x^m=(x^0,x^1,x^2,x^3)$.   We take even fields
to be continuous across the brane.  Odd fields, in general, can be
discontinuous across $\Si$, with a jump that is twice their value on either
side of the brane.  We make the effects of this discontinuity explicit by
redefining all odd fields and parameters as follows,
\bea
\label{redef}
\bma
e_m^\hf  \rightarrow \ep(z) \, e_m^\hf, &
e_5^a \rightarrow \ep(z) \, e_5^a, &
B_m \rightarrow \ep(z) \, B_m \\[2mm]
\eta_2 \rightarrow \ep(z) \, \eta_2, &
\psi_{m2} \rightarrow \ep(z) \, \psi_{m2}, &
\psi_{51} \rightarrow  \ep(z) \, \psi_{51} \\[2mm]
{} & q_3  \rightarrow \ep(z) \, q_3, &{}\\
\ema
\eea
where $\ep(z)$ is the sign function on $\mathbb{R}$.  After this redefinition,
all fields are {\it even}; the odd parity arises because of the $\ep(z)$ terms.
In what follows, we write all expressions in terms of the redefined fields.

The discontinuities in the fields are induced by the brane action.  We take
\bea
\label{brac}
S_{\rm brane} &=& \int_\Si \!d^4x e_4 \,   \left[
-3\la_1 -2 \alpha_1 \psi_{m1}\si^{mn}\psi_{n1} + h.c. \right] \nn\\
&=& \int\!d^5\!x e_4 \, \left[
-3\la_1 -2 \alpha_1 \psi_{m1}\si^{mn}\psi_{n1} + h.c. \right] \dz .
\eea
The bosonic piece describes the brane tension, $T_1=6\la_1$.  The fermionic
term is necessary to supersymmetrize the full bulk-plus-brane system for
arbitrary $\vec q$.

The bulk-plus-brane equations of motion can be readily computed.  In terms of
the redefined variables, the bosonic equations contain terms proportional to
$\delta'(z)$ and $\delta(z)^2$.  We eliminate these terms by demanding that all
odd bosonic fields vanish on the brane,
\begin{equation}
\label{simpl}
e_m^\hf= e_5^a = B_m = 0,
\end{equation}
on $\Si$. (This implies that $e_m^a$ on $\Si$ is the induced veirbein, 
so $d^4\!x e_4$ is the invariant integration measure on the brane.)
The equations of motion for $e_m^a$ and $\psi_{m1}$ contain terms proportional
to $\delta(z)$.
These terms cancel when
\bea
\label{bcom}
\om_{ma\hf}&=&\ep(z) \la_1 e_{ma} \\
\label{bcpm}
\psi_{m2}&=&\alpha_1 \psi_{m1},
\eea
on $\Si$.  The first condition restricts the spin connection on $\Si$; it is a
Neumann boundary condition for the metric $g_{mn}$.  The second condition
identifies the two gravitini on the brane.

\subsection{Supersymmetry transformations}

Consistency of the bulk-plus-brane theory requires closure of the supersymmetry
algebra and preservation of the boundary conditions on $\Si$.  To check the
closure, we convert the transformations (\ref{bulksusy}) to two-component
notation and carry out the redefinition (\ref{redef}) (see Appendix B).  It is
not hard to show that the algebra closes, except for the following singular
term,
\begin{equation}
[\da_\xi, \, \da_\eta]e_5^a=\dots{}+8i\eta_2\si^a\ov\xi_2\da(z) + h.c.
\end{equation}
We cancel this term by modifying the transformation for $\psi_{52}$:
\begin{equation}
\da\psi_{52}= \da\psi_{52}\Big\vert_{\rm old} - 4 \eta_2\da(z) .
\label{modsusy}
\end{equation}
This modification makes $\da\psi_{52}$ finite on $\Si$ and restores the
consistency of the supersymmetry algebra.

Supersymmetry also requires that the boundary conditions (\ref{simpl}) be
preserved:
\begin{equation}
\da e_m^\hf = \da e_5^a = \da B_m = 0 ,
\end{equation}
on $\Si$.  This imposes additional boundary conditions on the fermionic
fields:
\bea
\label{bceta}
\eta_2&=&\aone\eta_1 \\
\label{bcp5}
\psi_{51}&=&-\aone^\ast\psi_{52} ,
\eea
on $\Si$.

The boundary conditions (\ref{bcom}), (\ref{bcpm}) and (\ref{bcp5}) must
themselves be maintained under supersymmetry.  Equation (\ref{bcpm}) requires
the vanishing on $\Si$ of 
\begin{equation}
\label{vanM}
\aone\da\psi_{m1}-\da\psi_{m2}=-2\eta_1 \p_m\aone +iM\si_m \etabar_1,
\end{equation}
where
\begin{equation}
\label{M}
M=\la_1(1+\aone\aone^\ast \ep^2)+\la\left(
\aone\qstar+\aone^\ast q_{12}+(\aone\aone^\ast \ep^2-1)q_3\right)
\end{equation}
and $q_{12} = q_1 + i q_2$.  Equation (\ref{vanM}) implies that $\aone$ is
constant on $\Si$.  It also implies $M=0$, which gives the following relation
between $\la_1$ and $\aone$:
\begin{equation}
\label{cond1}
\la_1(1+\aone\aone^\ast)
+\la\left(\aone\qstar+\aone^\ast q_{12}
+(\aone\aone^\ast-1)q_3\right) = 0,
\end{equation}
where we have used the fact that $\ep^2=1$ (on $\Si$ and in the bulk).
The variations of eqs.~(\ref{bcp5}) and (\ref{bcom})
give boundary conditions on $\p_5\eta_2$ and $\p_5\psi_{m2}$, respectively.

We now have what we need to check the invariance of the bulk-plus-brane action
under supersymmetry.  The total variation receives three contributions:  two
from the bulk and one from the brane.  The first contribution comes from the
redefinition $q_3\rightarrow \ep(z) q_3$ in the bulk action:
\begin{equation}
\label{da1}
\da^{(1)} S_5 = \int\!d^5\!x e_4  \left[
6\la q_3 \left(\aone\aone^\ast \ep^2 -1 \right)
i\psi_{m1}\si^m\etabar_1 + h.c. \right] \dz.
\end{equation}
The second contribution arises from the modification (\ref{modsusy}) of the
supersymmetry transformation:
\bea
\label{da3}
&&\da^{(2)} S_5 = \int d^5\!x e_4 \Big[
-4\aone\eta_1\si^{mn}\Dh_m\psi_{n1}
+4\aone\psi_{n1}\si^{nm}\Dh_m\eta_1
-\sx i\aone\effh F^{m5}\psi_{m1}\eta_1
\nn\\
&&\hspace{2cm}
+6\left( \la_1\aone\aone^\ast\ep^2
+\la(\aone^\ast q_{12} +\aone\aone^\ast\ep^2 q_3)\right)
i\psi_{m1}\si^m\etabar_1
+ h.c. \Big] \dz\; .
\eea
The third contribution comes from the variation of the brane action.  The
supersymmetry transformations are induced from the bulk,
\bea
\da e_m^a &=& i \left(1+\aone\aone^\ast\ep^2 \right)
\eta_1\si^a\psibar_{m1} + h.c. 
\nn\\[2mm]
\da \psi_{n1} &=& 2\Dh_n\eta_1 +i \left(
\la\left( \qstar +\aone^\ast\ep^2 q_3 \right)
+\la_1\aone^\ast\ep^2 \right)\si_n\etabar_1
\nn\\
&&
-\frac{2}{\sx} i\,\effh F^{k5}\left(\si_{nk}+g_{nk}\right)\eta_1.
\eea
The variation of the brane action is then
\bea
\label{da2}
\da S_B &= &\int\!d^5\!x e_4  \big[
-8\aone\psi_{n1}\si^{nm}\Dh_m\eta_1
+\sx i\aone\effh F^{m5}\psi_{m1}\eta_1
\nn\\
&& + 6\left(
\la_1(1+2\aone\aone^\ast\ep^2)
+\la(\aone\qstar+\aone\aone^\ast\ep^2 q_3)
\right) i\psi_{m1}\si^m\etabar_1 + h.c. \big] \dz.
\eea
In these expressions, we have used the boundary conditions for the fermions and
the spin connection $\om_{ma\hf}$.

The supersymmetry variation of the bulk-plus-brane action is the sum of
(\ref{da1}), (\ref{da3}) and (\ref{da2}),
\begin{equation}
\label{datot}
\da (S_5+S_B) = \int d^5\!x e_4  \Big[
\Dh_m(-4\aone\psi_{n1}\si^{nm}\eta_1)
+6i \wt M \psi_{m1}\si^m\etabar_1 + h.c. \Big] \dz , 
\end{equation}
where
\begin{equation}
\label{mtilde}
\wt M = \la_1 \left(1+3\aone\aone^\ast\ep^2 \right)
+\la\left(\aone\qstar+\aone^\ast q_{12} 
+(3\aone\aone^\ast\ep^2 -1)q_3 \right).
\end{equation}
The derivative term in eq.~(\ref{datot}) integrates to zero because the
hatted derivative, defined in (\ref{dhat}), reduces to the covariant derivative
on $\Si$.  The other term vanishes, $\wt M\, \da(z)=0$, because of (\ref{cond1})
and the fact that $\ep^2 \da(z) = \frac{1}{3} \da(z)$.  Therefore the full
bulk-plus-brane action is supersymmetric, without any further conditions.

Equation (\ref{cond1}) defines the brane tension $\la_1$ in terms of
$\alpha_1$,
\begin{equation}
\la_1 = -\frac{\aone\qstar+\aone^\ast q_{12}+(\aone\aone^\ast-1)q_3}
{1+\aone\aone^\ast}\la .
\end{equation}
Using the (complex) Cauchy--Buniakovsky--Schwarz inequality,
$|\vec a \cdot \vec b|^2 \leq 
|\vec a|^2 \,
|\vec b|^2$,
we find
\begin{equation}
\label{ineq}
\left(\frac{\la_1}{\la}\right)^2 \leq \qq =1.
\end{equation}
This equation places an upper limit on the absolute value of the brane tension.

\subsection{Two branes}

This construction can be readily generalized to include a second brane.  We now
take the $x^5$ direction to have the topology of a circle $S^1$, and use the
following parametrization for the fifth dimension,
\begin{equation}
\label{defs1}
S^1 = [-z_2,\; -z_1]\cup[z_1,\; z_2], \quad
\ep(z)=\left\{
\bma
+1, & z\in S^1_+ \equiv &(z_1,\; z_2)\\
-1, & z\in S^1_- \equiv &(-z_2,\; -z_1),
\ema
\right.
\end{equation}
where we identify $-z_1\equiv z_1$ and $-z_2\equiv z_2$.  With these
definitions, $\ep^\pr(z)$ changes to
\begin{equation}
\ep^\pr(z) = 2[\da(z-z_1)-\da(z-z_2)] \equiv 2[\da_1(z)-\da_2(z)] .
\end{equation}
The parity operation identifies $z$ with $-z$.  It gives rise to two fixed
points, located at $z_1$ and $z_2$.  As before, we place 3-branes at the fixed
points.  We take all fields and parameters to have the parity assignments
(\ref{paras}), so their values in $z\in S^1_-$ are completely determined by
those in $z\in S^1_+$.   We work with fields redefined following
(\ref{redef}).

We construct the supersymmetric bulk-plus-brane action by
\begin{enumerate}
\item
introducing independent brane actions at the fixed points $\Sigma_1$ and
$\Sigma_2$,
\bea
S_B &=& \int_{\Si_1} d^4\!x e_4 \,
[-3\la_1-2\aone\psi_{m1}\si^{mn}\psi_{n1}+h.c.]
\nn\\
&& -\int_{\Si_2} d^4\!x e_4 \,
[-3\la_2-2\al_2\psi_{m1}\si^{mn}\psi_{n1}+h.c.] \\
&=& \int d^5\!x e_4 \left[
-3(\la_1\da_1(z)-\la_2\da_2(z)) 
-2(\alpha_1\da_1(z)-\alpha_2\da_2(z))\psi_{m1}\si^{mn}\psi_{n1} 
+ h.c. \right] ,\nn
\eea
\item
modifying the supersymmetry transformations,
\begin{equation}
\da\psi_{52} =\da\psi_{52}\Big\vert_{\rm old}
-4(\alpha_1\da_1(z)-\alpha_2\da_2(z))\eta_1 ,
\end{equation}
\item
and imposing the following boundary conditions\footnote{The boundary conditions
are determined by the parity assignments and the jump conditions that follow
from the brane action. We could have worked on the interval, instead of the covering
space $S^1$ with the orbifold identifications. These boundary conditions would have
then guaranteed vanishing of the boundary term of supersymmetry variation for the
bulk action with arbitrary $\vec q$.}
on $\Si_{1,2}$:
\bea
\label{2bc}
&&e_m^\hf = e_5^a = B_m = 0, \quad
\om_{ma\hf}=\ep(z)\la_{1,2} e_{ma} 
\nn\\[2mm]
&&\eta_2 =\alpha_{1,2}\eta_1, \quad
\psi_{m2}=\alpha_{1,2}\psi_{m1}, \quad
\psi_{51}= -\alpha_{1,2}^\ast\psi_{52} ,
\eea
\end{enumerate}
where $\alpha_{1,2} \in \mathbb{C}$ and $\la_{1,2} \in \mathbb{R}$ are
constants, related as follows,\footnote{
The results of \cite{abn} and
\cite{gp,flp,bkvp}  follow from ours if one sets $\vec q =(-1,0,0)$,
$\aone=\al_2=1$ and $\vec q =(0,0,1)$, $\aone=\al_2=0$, respectively.  In each
case $\la_1=\la_2 = \la$, which corresponds to the original Randall-Sundrum
scenario with opposite-tension branes. The case with $|\la_1|=|\la_2|\leq\la$
and $\al_1=\al_2=0$ was discussed in ref.~\cite{bfl}.}

\begin{equation}
\label{la12}
\la_{1,2} = -\frac{
\alpha_{1,2}\qstar+\alpha_{1,2}^\ast q_{12}
+(\alpha_{1,2}\alpha_{1,2}^\ast-1)q_3}
{1+\alpha_{1,2}\alpha_{1,2}^\ast}\la .
\end{equation}
The brane tensions $T_1=6\la_1$ and $T_2=-6\la_2$ are bounded by the
inequality
\begin{equation}
\label{l12ineq}
|\la_{1,2}| \leq \la .
\end{equation}
In the next section we will see that the bulk-plus-brane system has a
consistent dimensional reduction down to four dimensions.  The resulting
effective theory is $N=1,\ D=4$ (on-shell) supergravity with zero or negative
cosmological constant.

\section{Effective action}

In this section we derive the effective action for the supergravity zero modes
$e_m^a$ and $\psi_m$.  For simplicity, we ignore the radion multiplet and set
the radion field at its expectation value.  The zero modes for $e_m^\hf$,
$e_5^a$ and $B_m$ vanish because of the boundary condition (\ref{simpl}).

For the following, we restore the gravitational coupling $k_5$ by rescaling the
action and fields as
\begin{equation}
S \rightarrow k_5^2 S, \quad
B_M \rightarrow k_5 B_M, \quad
\psi_M \rightarrow k_5 \psi_M .
\end{equation}

\subsection{Bosonic reduction}

We first carry out the dimensional reduction for the bosonic part of the
action.  We take our ansatz to be
\begin{equation}
\label{ban}
e_m^a(x,z)=a(z)\wh e_m^a(x), \quad
e_m^\hf = e_5^a =0, \quad \effh =1, \quad B_m=B_5=0 .
\end{equation}
With this ansatz, the five-dimensional interval is
\begin{equation}
ds^2 =g_{MN}dx^M dx^N=a^2(z)\wh g_{mn}(x)dx^m dx^n+dz^2,
\end{equation}
and the connection coefficients are
\begin{equation}
\om_{mab}=\wh\omega_{mab}, \quad
\om_{ma\hf}=-a^\prime(z) \wh e_{ma}, \quad
\om_{5ab}=\om_{5a5}=0 ,
\end{equation}
where $\wh\omega_{mab}$ is the four-dimensional connection for $\wh e_m^a$.

Since $B_M=0$, the bosonic part of the bulk-plus-brane action is
\begin{equation}
\label{bac}
k_5^2 S_B = \int\!d^5\!x e_5 \left(-\frac{1}{2}R -\Lambda_5\right)
-\int_{\Si_1}\!d^4x e_4 T_1 
-\int_{\Si_2}\!d^4x e_4 T_2,
\end{equation}
where $\Lambda_5 = - 6\la^2$, $T_1=6\la_1$, and $T_2=-6\la_2$.  The
five-dimensional Einstein equations are
\begin{equation}
G_{mn}=\big(\Lambda_5+T_1\da_1(z)+T_2\da_2(z)\big)g_{mn}, \quad
G_{m5} = 0 , \quad
G_{55} =\Lambda_5,
\end{equation}
where  $G_{MN}=R_{MN}-\half R g_{MN}$ is the five-dimensional Einstein tensor.

The $G_{m5}=0$ equation is trivially satisfied for our ansatz.  The other two
equations reduce to
\bea
\label{beq1}
\wh G_{mn} &=& \Big[
3(a a^{\pr\pr} +{a^\pr}^2)- 6\la^2 a^2 + 6(\la_1\da_1(z)-\la_2\da_2(z))a^2\Big]
\wh g_{mn}\\[2mm]
\label{beq2}
\wh R &=& -12({a^\pr}^2 - \la^2 a^2) ,
\eea
where $\wh G_{mn}$ and $\wh R$ are the four-dimensional Einstein tensor and
scalar curvature for the metric $\wh g_{mn}$.

Using separation of variables, we split eq.~(\ref{beq1}) into
\begin{equation}
\label{b4eq}
\wh G_{mn}=\Lambda_4\wh g_{mn},
\end{equation}
and 
\begin{equation}
\label{beq3}
3(a a^{\pr\pr} +{a^\pr}^2) - 6\la^2 a^2 + 6(\la_1\da_1(z)-\la_2\da_2(z))a^2
=\Lambda_4,
\end{equation}
which are equations for $\wh g_{mn}(x)$ and $a(z)$, respectively. 
The separation constant $\Lambda_4$ is the cosmological constant in four
dimensions, according to equation (\ref{b4eq}).  We will soon find that
supersymmetry requires $\Lambda_4\leq 0$. For this case we write
\begin{equation}
\label{defK}
\Lambda_4= - 3 \la^2 K^2.
\end{equation}
(For the bosonic reduction alone, the case with positive cosmological constant
is obtained by replacing $K^2 \rightarrow - K^2$ here and in eq.~(\ref{a0eq})
below.)

The delta functions in eq.~(\ref{beq3}) must be cancelled by corresponding
singularities of $a^{\pr\pr}(z)$. Since $a(z)$ is an even function on
$S^1/\mathbb{Z}_2$, we write
\begin{equation}
\label{a0y}
a(z)=a_0(y), \qquad  y=\ep(z)\la z=\la |z|,
\end{equation}
where $a_0(y)$ is a smooth function on $\mathbb{R}$.  The derivatives of
$a_0(y)$ are well-defined, and we have
\begin{equation}
a^{\pr\pr}(z) =\la^2 a_0^{\pr\pr}(y)+2\la(\da_1(z)-\da_2(z))a_0^\pr(y).
\end{equation}
With these redefinitions, eqs.~(\ref{beq1}) and (\ref{beq2}) are simply
\begin{equation}
\label{a0eq}
a_0^{\pr\pr} = a_0, \quad
{a_0^\pr}^2 = a_0^2 - K^2,
\end{equation}
with boundary conditions
\begin{equation}
\label{a0bc}
\la_{1,2}=-\la\frac{a_0^\pr}{a_0}(y_{1,2}).
\end{equation}
The latter follow from the boundary conditions for the connection
$\om_{ma\hf}$.

We now proceed to find the effective action, without explicitly solving for the
warp factor $a_0$.  Indeed, using (\ref{a0eq}) and (\ref{a0bc}), together with
\begin{equation}
R=a^{-2}\big(\wh R +8a a^{\pr\pr}+12{a^\pr}^2\big) ,
\end{equation}
and
\begin{equation}
\oint dz a_0^3 a_0^\pr(\da_1(z)-\da_2(z)) =-\frac{\la}{2}
\oint dz(a_0^3 a_0^\pr)^\pr,
\end{equation}
we cast (\ref{bac}) in the form
\begin{equation}
\label{b4ac}
S_B = \frac{1}{k_5^2}\oint\!dz a_0^2 \int\!d^4\!x \wh e_4
\Big(-\frac{1}{2}\wh R - \Lambda_4\Big),
\end{equation}
where $\wh e_4=\text{det}(\wh e_m^a)=\sqrt{-{\rm{det}}(\wh g_{mn})}$.  If we
define the effective four-dimensional gravitational coupling to be
\begin{equation}
\label{keff}
\frac{1}{k_4^2}=\frac{1}{k_5^2}\oint dz a_0^2,
\end{equation}
we recover the action for four-dimensional gravity with a cosmological constant
$\Lambda_4$,
\begin{equation}
S_B = \frac{1}{k_4^2} \int\!d^4\!x \wh e_4
\Big(-\frac{1}{2}\wh R - \Lambda_4\Big).
\end{equation}
Equations (\ref{b4eq}) follow from this action, so the dimensional reduction is
consistent.

It is easy to solve eqs.~(\ref{a0eq}) to find the explicit form for the bosonic
warp factor \cite{kal}:
\begin{equation}
\label{adm}
\begin{array}{lllc}
\Lambda_4=-3\la^2 K^2&AdS_5\rightarrow AdS_4 & 
   a_0(y)=K \cosh(y-y_0)&|\la_{1,2}|<\la \\
\Lambda_4=0&AdS_5\rightarrow Mink_4 & 
   a_0(y)=\exp(\pm (y-y_0))& \la_1=\la_2=\mp\la\\
\Lambda_4=+3\la^2 K^2&AdS_5\rightarrow dS_4 & 
   a_0(y)=K \sinh(y-y_0)&|\la_{1,2}|>\la .
\end{array}
\end{equation}
(We use $dS/AdS/Mink$ to denote a theory with positive/negative/zero
cosmological constant.)
The restrictions on $\la_{1,2}$ follow from boundary conditions (\ref{a0bc})
and the fact that $|\tanh(y)|<1$.  The case $\Lambda_4=0$ corresponds to the usual
Randall-Sundrum scenario with two opposite-tension branes.

In the previous section we found that local supersymmetry places a restriction
(\ref{l12ineq}) on the brane tensions: $|\la_{1,2}|\leq \la$.  This implies,
according to (\ref{adm}), that local supersymmetry restricts the effective
four-dimensional theory to be $AdS_4$ or $Mink_4$.  Note that for $Mink_4$, the
parameter $y_0$ in (\ref{adm}) is arbitrary and interbrane distance is not
fixed.  For $AdS_4$ the boundary conditions (\ref{a0bc}) imply $y_0 =
y_1+\rm{Arctanh}(\la_1/\la)$ and fix the proper distance $\Delta z$ in
terms of $\la_1,\,\la_2$ and $\la$:
\begin{equation}
\label{delz}
\la \Delta z = \text{Arctanh}(\frac{\la_1}{\la})
-\text{Arctanh}(\frac{\la_2}{\la})
=\frac{1}{2}\ln\left[
\frac{(\la+\la_1)(\la-\la_2)}
     {(\la+\la_2)(\la-\la_1)}\right] .
\end{equation}

The effective cosmological constant $\Lambda_4$ is determined once we normalize
the bosonic warp factor. It is natural to require that $a_0(y)$ is unity at the
location of an observer (so that s/he uses the same time and distance scales
for both the five-dimensional and effective four-dimensional theory). When
effective theory is $AdS_4$, we use $\cosh(y)\geq1$ to find that $0<K\leq 1$ and
\begin{equation}
-3\la^2\leq\Lambda_4<0 .
\end{equation}

\subsection{Fermionic reduction}

The fermionic part of the four-dimensional effective action is fixed by
supersymmetry.   The five-dimensional gravitini must reduce to a
four-dimensional gravitino $\psi_m(x)$, while the five-dimensional
supersymmetry parameters must reduce to a four-dimensional spinor $\eta(x)$.
These considerations motivate the following ansatz for the fermionic fields,
\begin{equation}
\label{fan}
\begin{array}{lll}
\eta_1(x,z)=\beta_1(y)\eta(x),
&\psi_{m1}(x,z)=\gamma\beta_1(y)\psi_m(x)
&\psi_{51}(x,z)=0,\\
\eta_2(x,z)=\beta_2(y)\eta(x),
&\psi_{m2}(x,z)=\gamma\beta_2(y)\psi_m(x),
&\psi_{52}(x,z)=0,
\end{array}
\end{equation}
where the complex warp factors are functions of $y=\la|z|$, just like $a_0$ is
a function of $y$ in eq.~(\ref{a0y}).

Supersymmetry imposes the following consistency conditions on this ansatz:
\begin{equation}
\beta_1\da\psi_{m2}=\beta_2\da\psi_{m1}, \quad
\da\psi_{51}=0, \quad \da\psi_{52}=0.
\end{equation}
The first condition requires
\begin{equation}
\label{bbaq}
(\beta_1\beta_1^\ast+\beta_2\beta_2^\ast)\frac{a_0^\pr}{a_0} =
q_{12}\beta_1\beta_2^\ast+q_{12}^\ast\beta_1^\ast\beta_2 +
q_3(\beta_2\beta_2^\ast-\beta_1\beta_1^\ast).
\end{equation}
This equation, restricted to the fixed points, together with boundary
conditions
\begin{equation}
\label{bcalbb}
\al_{1,2}= \frac{\beta_2}{\beta_1}(y_{1,2})
\end{equation}
and (\ref{a0bc}), implies the relation (\ref{la12}) between $\la_{1,2}$ and
$\al_{1,2}$.
The other two conditions, $\da\psi_{51}=0$ and $\da\psi_{52}=0$, give rise to
the following equations,
\begin{equation}
\label{b12pr}
2\beta_1^\pr = \qstar\beta_2 - q_3\beta_1, \quad
2\beta_2^\pr = q_{12}\beta_1 + q_3\beta_2 .
\end{equation}
They permit the right-hand side of eq.~(\ref{bbaq}) to be written as
$(\beta_1\beta_1^\ast+ \beta_2\beta_2^\ast)^\pr$, which implies that the
bosonic and fermionic warp factors obey
\begin{equation}
\label{a0bb}
a_0 = \beta_1\beta_1^\ast+\beta_2\beta_2^\ast ,
\end{equation}
up to a multiplicative constant that we set equal to unity.

These conditions are sufficient to find the supersymmetry transformations of
the four-dimensional fields.  Using the five-dimensional transformation $\da
e_m^a$, together with the combination $\beta_1^\ast\da\psi_{m1}
+\beta_2^\ast\da\psi_{m2}$, we find
\bea
\label{dagamma}
\da \wh e_m^a &=& i k_5\gamma^\ast \eta\si^a\psibar_m + h.c. \nn\\[2mm]
k_5\gamma\da\psi_m &=& 2\Dh_m\eta +i \la g \wh\si_m\etabar,
\eea
where
\begin{equation}
\label{defg}
g^\ast = q_{12}\beta_1^2-\qstar\beta_2^2+2q_3\beta_1\beta_2 .
\end{equation}
Equations (\ref{b12pr}) and (\ref{a0bb}) assure us that $g$ is a constant, and
relate it to $K$ as follows,
\begin{equation}
gg^\ast = a_0^2 - {a_0^\pr}^2 = K^2 .
\end{equation}
The four-dimensional supersymmetry transformations (\ref{dagamma}) take their
usual form if we set
\begin{equation}
\label{gamkk}
\gamma = \frac{k_4}{k_5}.
\end{equation}

To find the effective action, we first note that the
$\psi_{m1}\si^{mn}\psi_{n2}\da_{1,2}(z)$ terms from the bulk action exactly
cancel the $\al_{1,2}\psi_{m1}\si^{mn}\psi_{n1}\da_{1,2}(z)$ terms from the
branes.  Therefore the fermionic part of the bulk-plus-brane action reduces to
\bea
S_F &=& \int d^5\!x e_4 
\Big\{
\frac{1}{2} \eps^{mpnk}\Big(
\psibar_{m1}\sibar_p D_n\psi_{k1}
+ \psibar_{m2}\sibar_p D_n\psi_{k2}\Big) \nn\\[2mm]
 &&\qquad\qquad +\ep(z)\Big(
\psi_{m1}\si^{mn}\p_5\psi_{n2}-\psi_{m2}\si^{mn}\p_5\psi_{n1}\Big)
\nn\\[2mm]
 &&\quad -\frac{3\la}{2}\Big(
q_{12}\psi_{m1}\si^{mn}\psi_{n1}-\qstar\psi_{m2}\si^{mn}\psi_{n2}
+2q_3\psi_{m1}\si^{mn}\psi_{n2}\Big) + h.c.
\Big\}.
\eea
Using eqs.~(\ref{b12pr}), (\ref{a0bb}) and (\ref{defg}), as well as
\begin{equation}
\si_m=a_0\wh\si_m, \quad
\si^{mn}=a_0^{-2}\wh\si^{mn}, \quad
\eps^{mpnk}=a_0^{-4}\wh\eps^{mpnk}, \quad
e_4=a_0^4\wh e_4 ,
\end{equation}
we transform this action to
\begin{equation}
S_F = \gamma^2 
\oint\!dz a_0^2\int\!d^4\!x \wh e_4 \left[
\frac{1}{2}\wh\eps^{mpnk}\psibar_m\wh\sibar_p\Dh_n\psi_k
-\la g^\ast \psi_m\wh\si^{mn}\psi_n +h.c. \right] .
\end{equation}
The normalization conditions (\ref{keff}) and (\ref{gamkk}) ensure that
$\gamma^2 \oint\!dz a_0^2 = 1$. Together with the bosonic part, the total
effective action is therefore
\begin{equation}
S_4 = \int\!d^4\!x \wh e_4 \left\{ 
\frac{1}{k_4^2}\left(-\frac{1}{2}\wh R +3\la^2 g g^\ast\right)
+ \left[
\frac{1}{2}\wh\eps^{mpnk}\psibar_m\wh\sibar_p\Dh_n\psi_k
-\la g^\ast \psi_m\wh\si^{mn}\psi_n +h.c. \right] \right\} ,
\end{equation}
which is the correct action for locally $N=1$ supersymmetric theory in four
dimensions. This completes the dimensional reduction.

If we restrict $\wh g_{mn}(x)$ to be the metric for the maximally symmetric
anti-de
Sitter (or Minkowski) background, the local supersymmetry breaks to 
$N=1$ global supersymmetry.  The unbroken supersymmetry is
described by the five-dimensional spinors
$\eta_{1,2}=\beta_{1,2}(y)\eta(x)$, where $\eta(x)$ is now fixed to be the 
four-dimensional Killing spinor in the $\wh g_{mn}(x)$ background
\cite{brfr}.

We also present here explicit solutions for the fermionic warp factors.
Equations (\ref{b12pr}) are straightforward to solve.  We first note that
$4\beta_{1,2}^{\pr\pr} = \beta_{1,2}$ and, using the boundary conditions
(\ref{bcalbb}), we obtain the following expressions,
\bea
\label{b1b2}
\beta_1(y)&=&b_0 \cosh\half(y-y_1)+(\qstar\al_1-q_3) b_0 \sinh\half(y-y_1) \\
\beta_2(y)&=&
   \al_1b_0\cosh\half(y-y_1)+(q_{12}+\al_1q_3)b_0\sinh\half(y-y_1).
\eea
The overall constant $b_0$ is fixed (up to a phase) by the normalization of
the bosonic warp factor.  Substituting these expressions into (\ref{defg}), we
find
\begin{equation}
g^\ast = b_0^2(q_{12}-\qstar\al_1^2+2q_3\al_1).
\end{equation}

We distinguish two cases.  If $\al_1$ is a root of 
\begin{equation}
\label{qroot}
q_{12}-\qstar\al_1^2+2q_3\al_1 = 0,
\end{equation}
the effective theory is $Mink_4$. This equation implies $\qstar\al_1-q_3=\pm1$
and $q_{12}+q_3\al_1=\pm\al_1$, which permits us to write
\begin{equation}
\beta_1(y)=b_0 \exp(\pm\half y), \quad
\beta_2(y)=\al_1\beta_1(y), \quad
a_0(y)=\displaystyle\frac{2 b_0 b_0^\ast}{1\mp q_3}\exp(\pm y).
\end{equation}
The boundary conditions (\ref{a0bc}) and (\ref{bcalbb}) require $\al_2=\al_1$
and $\la_2=\la_1=\mp\la$, independent of the interbrane distance.

For any complex $\al_1$ which is not a solution of (\ref{qroot}), the effective
theory is $AdS_4$.  The value of $\la_1$ is determined by $\al_1$,
\begin{equation}
\label{l1a1}
\la_1 = -\frac{\aone\qstar+\aone^\ast q_{12}+(\aone\aone^\ast-1)q_3}
{1+\aone\aone^\ast}\la,
\end{equation}
so $|\la_1|<\la$.  We introduce the real variable $\wh y_1=
\rm{Arctanh}(\la_1/\la)$ and, using (\ref{a0bb}) and (\ref{b1b2}), cast the
bosonic warp factor in the following form,
\begin{equation}
\label{a0K}
a_0(y) = K \cosh(y-(y_1+\wh y_1)),
\end{equation}
where
\begin{equation}
K = |g| = \frac{b_0 b_0^\ast (1+\al_1\al_1^\ast)}{\cosh\wh y_1}.
\end{equation}
For a given separation of the branes, $\Delta y=y_2-y_1$, the boundary
conditions (\ref{a0bc}) and (\ref{bcalbb}) determine the values of $\al_2$ and
$\la_2$,
\begin{equation}
\al_2=\frac{\al_1+(q_{12}+\al_1 q_3)\tanh(\half\Delta y)}{
1+(\qstar\al_1-q_3)\tanh(\half\Delta y)}, \quad
\la_2=\la\; \frac{\la_1-\la\tanh\Delta y}{\la-\la_1\tanh\Delta y}.
\end{equation}
It is not hard to check that $\al_2$ and $\la_2$  are related by
eq.~(\ref{la12}) and furthermore, that $\al_2$ cannot be a solution of
(\ref{qroot}).  Alternatively, for a given $\al_1$ and $\al_2$, the interbrane
separation is 
\begin{equation}
\label{restrict}
\Delta y =
2{\rm Arctanh}\left(\frac{\al_2-\al_1}{q_{12}-\qstar\al_1\al_2
+q_3(\al_1+\al_2)}\right).
\end{equation}
This equation is equivalent to (\ref{delz}), since $\la_{1,2}$ are given by
eq.~(\ref{la12}).

The fact that the argument of Arctanh must be real and of absolute value less
than unity implies that $\al_2$ cannot be chosen completely independently of
$\al_1$.  For example, if $q_3=1$, this restricts $\al_1$ and $\al_2$ to have
the same complex phase,
\begin{equation}
\al_1=r_1 e^{i\theta}, \quad \al_2=r_2 e^{i\theta}.
\end{equation}
For given $\la_{1,2}$, the absolute values of $\al_{1,2}$ are determined as
\begin{equation}
\label{r12}
r_{1,2}=\cosh\wh y_{1,2} -\sinh\wh y_{1,2}, \quad
\wh y_{1,2}=\rm{Arctanh}(\frac{\la_{1,2}}{\la}).
\end{equation}
For $q_3\neq 1$, one can first rotate to $\vec q\;' = (0,0,1)$, as explained in
Appendix \ref{ap2}, then use the above result and rotate back, arriving at
\begin{equation}
\al_1=\frac{1-q_3+q_{12}r_1 e^{i\theta}}{
(1-q_3)r_1 e^{i\theta}-\qstar}, \quad
\al_2=\frac{1-q_3+q_{12}r_2 e^{i\theta}}{
(1-q_3)r_2 e^{i\theta}-\qstar},
\end{equation}
where $r_{1,2}$ are given by the same expressions, and $\theta$ is arbitrary.
The apparent singularity at $q_3=1$ is a consequence of trying to cover the
sphere $\qq=1$ using a single coordinate patch.

\section{Summary and conclusions}

In this paper we presented a general bulk-plus-brane action for the 
supersymmetric Randall-Sundrum scenario.  The bulk action is that of 
$N=2$ supergravity, compactified on a five-dimensional 
$S^1/\mathbb{Z}_2$ orbifold.  The brane action contains supergravity 
fields induced from the bulk.  The bulk gravitino mass depends on a 
vector $\vec q$, parametrizing a point on the sphere $S^2$.

We demonstrated that our bulk-plus-brane action has local $N=2$ 
supersymmetry, constrained by boundary conditions for the fields and 
supersymmetry parameters.  The boundary conditions are those implied by 
consistency of the five-dimensional equations of motion and supersymmetry
transformations.  For the action 
we considered, the brane tensions $T_1=6\la_1$ and $T_2=-6\la_2$ respect 
an upper limit, expressed in terms of the bulk cosmological constant, 
$\Lambda_5=-6\la^2$, as $|\la_{1,2}|\leq\la$.  

We also presented a consistent dimensional reduction for the 
bulk-plus-brane system.  We derived the action and the supersymmetry 
transformations in a background-independent way, without explicitly 
solving for the warp factors.  The effective action is that of minimal 
$N=1$ supergravity in four dimensions, with zero or negative 
cosmological constant.  The effective cosmological constant is zero if 
and only if $\la_1=\la_2=\pm\la$, which corresponds to the original 
Randall-Sundrum scenario with two opposite-tension branes.  
Our results show, however, that
the Randall-Sundrum fine-tuning is not a consequence of supersymmetry.
For all other $|\la_{1,2}|<\la$, we consistently obtain $N=1$-supersymmetric 
effective theory in a space with a negative cosmological constant, limited by
\begin{equation}
\frac{1}{2} \Lambda_5 \leq \Lambda_4 < 0.
\end{equation}

When $\Lambda_5$ is nonzero, the gravitino mass localized
on each brane is determined by the brane tension and the bulk
cosmological constant.  In contrast to flat space, supersymmetry
cannot be broken spontaneously by changing the brane masses,
as was done in \cite{bff}.  Spontaneous supersymmetry breaking for
warped geometries is discussed in \cite{bfl,pom,bb}.

{\bf Acknowledgements.}
We would like to thank  A.\ Falkowski, D.\ Nemeschansky, S.\ Pokorski and 
R.-J.\ Zhang for helpful discussions.
This work was supported in part by the U.S. National Science Foundation, 
grant NSF-PHY-9970781.

{\bf Note added.}
When $\Lambda_5\neq 0$, the $SU(2)$ automorphism symmetry of the bulk
action is broken to a $U(1)$ R-symmetry that depends on $\vec q$.
The transformation $\Psi_{mi}\rightarrow\Psi_{mi}^\prime=U_i{}^j\Psi_{mj}$
leaves the bulk action (\ref{bulk}) invariant for $U=\exp[i({\vec q}\cdot{\vec\sigma})\phi]$,
where $\phi\in\mathbb{R}$.  It is a symmetry of the full theory
if it also preserves the boundary conditions (\ref{2bc}).  This is the case precisely
when eq.~(\ref{qroot}) is satisfied, that is, when effective theory has $\Lambda_4=0$.
The $U(1)$ R-symmetry of the five-dimensional theory gives rise to a $U(1)$
R-symmetry of the effective theory, $\psi_m\rightarrow\psi^\pr_m=\exp(\mp i\phi)\psi_m$.
We thank A.\ Nelson for raising this question.

\vspace{10mm}

\appendix
\section{Conventions}
\label{ap1}

In this paper we adopt the following index conventions,
\begin{equation}
\bma
M,N,P,Q,K & \text{coordinate space} & M=\{m,5\} & 
m=\{0,1,2,3\}\\[1mm]
A,B,C,D,E & \text{tangent space} & A=\{a,\hat5\} & 
a=\{\hat0,\hat1,\hat2,\hat3\}\\[1mm]
i,j & SU(2) & i=\{1,2\}.& {}
\ema
\end{equation}
We denote the determinant of an $n$-bein by $e_n$:
\begin{equation}
e_5={\rm det}e_M^A, \quad e_4={\rm det}e_m^a, \quad \wh e_4={\rm det}\wh e_m^a
.
\end{equation}
The f\"unfbein $e_M^A$ (and the veirbein $e_m^a$) allow one to convert between
the two types of indices,
\bea
\Ga_M = e_M^A \Ga_A, \quad
g_{MN} = e_M^A e_N^B \eta_{AB}, \quad
\eps^{MNPQK} = e_A^M e_B^N e_C^P e_D^Q e_E^K \eps^{ABCDE} .
\eea
The gamma matrices obey the following relations,
\bea
&&\left\{ \Ga_A, \; \Ga_B \right\} = -2\eta_{AB}, \quad
\eta_{AB} = {\rm diag} (-  +  +  +  +)
\nn\\[2mm]
&&\Ga^{ABCDE} = -\eps^{ABCDE}, \quad
\eps^{\hat0\hat1\hat2\hat3\hat5} = +1, \quad
\eps^{abcd\hf} = \eps^{abcd}
\eea
and
\begin{equation}
\Ga^{ABCD} = \eps^{ABCDE}\Ga_E, \quad
\Ga^{ABC} = \eps^{ABCDE}\Si_{DE}, \quad
\Ga^{AB} = \half\left[\Ga^A, \; \Ga^B\right] = 2\Si^{AB}.
\end{equation}

The reduction to two-component notation \cite{wb} exploits the following
representation for the gamma matrices,
\bea
\Ga^a = \bpm 0&\si^a \\ \sibar^a&0 \epm, \;
\Ga^\hf = \bpm -i&0 \\ 0&i \epm \;\Rightarrow\;
\Si^{ab} = \bpm \si^{ab} & 0 \\ 0 & \sibar^{ab} \epm, \;
\Si^{a\hf} = \frac{i}{2} \bpm 0 & \si^a \\ -\sibar^a & 0 \epm .
\eea
The charge conjugation matrix is taken to be
\bea
C = \bpm i\si_2 & 0 \\ 0 & i\si_2 \epm
  = \bpm -\eps_{\alpha\beta} & 0 \\ 0 & \eps^{\dot\alpha\dot\beta} \epm.
\eea
With this representation, a four-component Dirac spinor, its Dirac conjugate
and its Majorana conjugate are
\bea
\Psi = \binom{\psi_{1\alpha}}{\psibar_2^{\dot\alpha}}, \quad
\Psibar = \Psi^\dagger\Ga_{\hat0} 
=(\psi_2^\alpha , \; \psibar_{1\dot\alpha}), \quad
\Psitil = \Psi^T C = (-\psi_1^\alpha , \; \psibar_{2\dot\alpha}).
\eea
A symplectic Majorana spinor obeys the following condition
\begin{equation}
\Psitil^i = \Psibar_i .
\end{equation}
We take
\bea
\Psi_1 = -\Psi^2 = \binom{ \psi_1}{\psibar_2}, \quad
\Psi_2 =  \Psi^1 = \binom{-\psi_2}{\psibar_1} .
\eea

The covariant derivative and its commutator on a Dirac spinor are given by
\begin{equation}
D_M\Psi = \p_M\Psi +\half\om_{MAB}\Si^{AB}\Psi, \qquad
[D_M,  D_N]\Psi = \half R_{MNAB}\Si^{AB}\Psi .
\end{equation}
The connection coefficients, curvature tensor and scalar curvature are defined
as follows
\bea
&&\om_{MAB} = \half e_A^N e_B^K (e_{MC} \p_{[N}e_{K]}^C 
- e_{NC} \p_{[K}e_{M]}^C - e_{KC} \p_{[M}e_{N]}^C) \nn
\\[1mm]
&&R_{MNAB} =\p_M\om_{NAB}-\p_N\om_{MAB}
+\om_{NA}{}{}^C\om_{MCB} - \om_{MA}{}{}^C\om_{NCB} \nn
\\[2mm]
&&R = e^{MA}R_{MA} = e^{MA}e^{NB}R_{MNAB} .
\eea
We introduce a hatted covariant derivative by splitting $D_M$ as
\bea
\label{dhat}
&&D_M\Psi = \Dh_M\Psi + \om_{Ma\hf}\Si^{a\hf}\Psi
\quad\Longrightarrow\quad
\left\{\bma
D_M\psi_1 = \Dh_M\psi_1 +\frac{i}{2}\om_{Ma\hf}\si^a\psibar_2  \\
D_M\psi_2 = \Dh_M\psi_2 -\frac{i}{2}\om_{Ma\hf}\si^a\psibar_1  \\
\Dh_M\psi = \p_M\psi +\frac{1}{2}\om_{Mab}\si^{ab}\psi.
\ema\right.
\eea
When $e_m^\hf=e_5^a=0$, $\Dh_m$ becomes the covariant derivative for $e_m^a$.
When, in addition, $e_m^a=a(z)\wh e_m^a(x)$, $\Dh_m$ is also the covariant
derivative for $\wh e_m^a(x)$.

\section{The bulk supergravity action}
\label{ap2}

The action of pure $N=2,\,D=5$ supergravity without a cosmological constant is
invariant under the $SU(2)$ rotations $\Psi_{Mi}^\prime = \Uij\Psi_{Mj}$.  The
rotation $U \in SU(2)$ can be written in terms of the Pauli matrices,
\begin{equation}
\quad U^\dagger U = 1, \quad \det(U) = 1
\quad\Rightarrow\quad 
\Uij = u_0\si_0 + i\vec u \cdot\vec\si, \quad
u_0, u_i \in \mathbb{R}, \quad u_0^2+\vec u^{\,2} =1 .
\end{equation}
A cosmological constant is introduced by gauging a $U(1)$ subgroup of the
$SU(2)$.  The gauge coupling breaks the symmetry and changes the covariant
derivative $D_M\Psi_{Ni}$ into $D_M\Psi_{Ni} -\sqrt{\frac{3}{2}}\la B_M
\Qij\Psi_{Nj}$, where

\begin{equation}
\quad Q^\dagger = -Q, \quad Tr(Q) = 0
\quad\Rightarrow\quad
\Qij = i\vec q \cdot \vec\si=i\left(
\bma
q_3 & q_1-iq_2\\
q_1+iq_2 & -q_3
\ema\right), \quad
q_i \in \mathbb{R}.
\end{equation}
The matrix for the $SU(2)$ rotation on the two-component spinors,
$\psi_i^\prime = \wt U_i{}^j \psi_j$, is given by
\begin{equation}
\wt U =u_0\si_0+i(-u_1\si_1-u_2\si_2+u_3\si_3) =
\si_3 U \si_3 .
\end{equation}
Any such rotation can be compensated by changing $Q$
\bea
&&Q^\prime = U Q U^\dagger =
(u_0 +i\vec u \cdot \vec\si) i\vec q \cdot \vec\si (u_0 -i\vec u \cdot
\vec\si) =
i{\vec q}^{\;\prime} \cdot \vec\si
\nn\\
&& {\vec q}^{\;\prime} 
= \vec q +2\vec u\times(\vec u\times\vec q) -2u_0(\vec u\times\vec q) .
\eea

With these conventions, the action of gauged supergravity is\footnote{We assume
$\qq=1$.  This makes the cosmological constant $\Lambda_5=-6\la^2\qq=3\la^2
Tr(Q^2)$ independent of $\vec q$.  Our definition of $\Qij$ follows
ref.~\cite{bkvp}.}
\bea
S_5 &=& \int d^5\!x e_5 
\Big\{
-\frac{1}{2} R + 6\la^2\qq 
+\frac{i}{2}\Psitil_M^i \Ga^{MNK} D_N \Psi_{Ki}
+i\frac{3}{2}\la \Psitil_M^i \Si^{MN} Q_i{}^j \Psi_{Nj} 
\nn\\
&&\quad -\frac{1}{4}F_{MN}F^{MN} 
- i\frac{\sqrt6}{16}F_{MN}
\left(
2\Psitil^{Mi} \Psi_i^N + \Psitil_P^i \Ga^{MNPQ} \Psi_{Qi} 
\right) 
\nn\\
&&\quad -\frac{1}{6\sqrt6}\eps^{MNPQK} F_{MN}F_{PQ}B_K
-i\frac{\sqrt6}{4}\la B_N\Psitil_M^i\Ga^{MNK}\Qij\Psi_{Kj}
\Big\}.
\eea
For constant parameters, the Lagrangian is invariant (up to a total derivative)
under the following supersymmetry transformations
\bea
\da e_M^A &=& i \Etatil^i\Ga^A\Psi_{Mi} \nn\\[2mm]
\da B_M &=& i\frac{\sqrt6}{2}\Psitil_M^i\Eta_i \nn\\
\da \Psi_{Mi} &=& 2 \big(
D_M\Eta_i - \frac{\sqrt6}{2}\la B_M\Qij\Eta_j \big)
+\la\Ga_M\Qij\Eta_j
\nn\\
&&
+\frac{1}{2\sqrt6} \left(
\Ga_{MNK} - 4g_{MK}\Ga_N \right) F^{NK}\Eta_i .
\eea
If the parameters are not constant, as when we change constant $q_3$ by a
function $\ep(z)q_3$, the variation of the action is
\begin{equation}
\label{dilaqu}
\da S_5 = \int d^5\!x e_5 
\left\{
D_M \big(\dots\big)^M
-\left[6i\Psitil_M^i\Si^{MN}\Eta_j
+\sx i\Psitil_M^i\Ga^{MNK}\Eta_j B_K \right]
\p_N \big(\la \Qij)
\right\} .
\end{equation}

The action and transformation laws can also be written in terms of two
component spinors.  
We use the identities
\bea
\label{help}
&&i\Psitil^i\dvec\Ga\Eta_i=
i\Psibar_1\dvec\Ga\Eta_1 +h.c.
\nn\\
&& i\Psitil^i\dvec\Ga\Qij\Eta_j=
-q_3\Psibar_1\dvec\Ga\Eta_1 -q_{12}\Psibar_2\dvec\Ga\Eta_1 +h.c.\;,
\eea
where $q_{12}=q_1+iq_2$, $\dvec\Ga = \Ga^{A_1}\Ga^{A_2}\dots\Ga^{A_n}$, and
$\Psi^i$ and $\Eta^i$ are arbitrary symplectic Majorana spinors.  We carry out
the redefinition (\ref{redef}) and set $e_m^\hf = e_5^a = B_m = 0$. 
The fermionic part of the bulk action is then
\bea
\label{epac}
S_{5F} &=& \int d^5\!x e_5 \efhf 
\Big\{
\frac{1}{2}\eps^{mpnk}\big(
\ep^2\psibar_{m2}\sibar_p D_n\psi_{k2}
+ \psibar_{m1}\sibar_p D_n\psi_{k1}\big)\effh
\nn\\
&+& \big(\psi_{52}\si^{mn}D_m\psi_{n1}
      -\ep^2\psi_{51}\si^{mn}D_m\psi_{n2}\big)
+\big(\ep^2\psi_{m2}\si^{mn}D_n\psi_{51}
      -\psi_{m1}\si^{mn}D_n\psi_{52}\big)
\nn\\
&-& \frac{3\la}{2}\Big[
\ep^2 q_3 
\big( \left(\psi_{m2}\si^{mn}\psi_{n1}+\psi_{m1}\si^{mn}\psi_{n2}\right)\effh
+i\left(\psi_{m2}\si^m\psibar_{52}-\psi_{m1}\si^m\psibar_{51}
\right) \big)
\nn\\
&& + q_{12} \big( \left(
\psi_{m1}\si^{mn}\psi_{n1}
-\ep^2\psibar_{m2}\sibar^{mn}\psibar_{n2}\right)\effh
+i\left(\psi_{m1}\si^m\psibar_{52}
+\ep^2\psibar_{m2}\sibar^m\psi_{51}
\right) \big) \Big]
\nn\\
&-&\frac{\sqrt6}{4}i\effh F^{m5} \left(
\ep^2\psi_{m2}\psi_{51} -\psi_{m1}\psi_{52} \right)
+\frac{\sqrt6}{8}iF_{m5}\eps^{mnpq} \left(
\ep^2\psi_{p2}\si_n\psibar_{q2} +\psi_{p1}\si_n\psibar_{q1} \right)
\nn\\
&+& \!\frac{\sqrt6}{2}i\la B_5 \Big[
\ep^2 q_3 \left(
\psi_{m2}\si^{mn}\psi_{n1} +\psi_{m1}\si^{mn}\psi_{n2} \right)
+q_{12}\left(
\psi_{m1}\si^{mn}\psi_{n1} +\ep^2\psibar_{m2}\sibar^{mn}\psibar_{n2}
\right) \Big]
\nn\\
&-&\ep \big(\psi_{m2}\si^{mn}D_5\psi_{n1} 
          -\psi_{m1}\si^{mn}D_5\psi_{n2}\big)
+2\psi_{m1}\si^{mn}\psi_{n2}\dz + h.c. 
\Big\},
\eea
and the supersymmetry transformations are
\bea
\label{epsusy}
\da e_m^a &=& i \left(
\ep^2\eta_2\si^a\psibar_{m2} +\eta_1\si^a\psibar_{m1} 
\right) + h.c.
\nn\\
\da e_5^a &=& i \left(
\eta_2\si^a\psibar_{52} +\eta_1\si^a\psibar_{51} 
\right) + h.c. 
\nn\\
\da e_m^\hf &=& \eta_2\psi_{m1} - \eta_1\psi_{m2} + h.c.
\nn\\
\da \effh &=& \ep^2\eta_2\psi_{51} - \eta_1\psi_{52} + h.c.
\nn\\
\da B_m &=& i\frac{\sqrt6}{2} \left( 
    \psi_{m2}\eta_1 - \psi_{m1}\eta_2 \right) + h.c.
\nn\\
\da B_5 &=& i\frac{\sqrt6}{2} \left( 
    \psi_{52}\eta_1 - \ep^2\psi_{51}\eta_2 \right) + h.c.
\nn\\
\da \psi_{m1} &=& 2D_m\eta_1 + i\la\si_m \left(
   \ep^2 q_3\etabar_2 +\qstar\etabar_1 \right)
-\frac{2}{\sx} i\effh F^{n5} \left( \si_{mn} + g_{mn} \right) \eta_1
\nn\\
\da \psi_{m2} &=& 2D_m\eta_2 + i\la\si_m \left(
   q_3\etabar_1 -q_{12}\etabar_2 \right)
-\frac{2}{\sx} i\effh F^{n5} \left( \si_{mn} + g_{mn} \right) \eta_2
\nn\\
\da \psi_{51} &=& 2\ep^{-1}D_5\eta_1 
+\la(\effh-\sx i B_5)( q_3\eta_1 -\qstar\eta_2)
-\frac{2}{\sx}F_{m5}\si^m\etabar_2
\nn\\
\da \psi_{52} &=& 2\ep D_5\eta_2 
-\la(\effh-\sx i B_5)( \ep^2 q_3\eta_2 +q_{12}\eta_1)
+\frac{2}{\sx}F_{m5}\si^m\etabar_1
+ 4\eta_2\dz, \nn\\
\eea
where
\begin{equation}
D_m\eta_1=\Dh_m\eta_1+\frac{i}{2}\ep\om_{ma\hf}\si^a\etabar_2, \quad
D_m\eta_2=\Dh_m\eta_2-\frac{i}{2}\ep^{-1}\om_{ma\hf}\si^a\etabar_1 ,
\end{equation}
and similarly for other covariant derivatives, according to (\ref{dhat}).  (The
corresponding expressions \emph{before} the redefinition (\ref{redef}) are
obtained by setting $\ep=1$ and dropping the $\da(z)$ terms.)


\end{document}